\documentclass[12pt]{iopart}

\usepackage{graphicx}
\usepackage{dcolumn}
\usepackage{bm}

\usepackage{iopams}
\begin{document}

\title{Optics of spin-1 particles from gravity-induced phases}

\author{G. Papini$^{a,b,c}$, G. Scarpetta$^{c,d,e}$, A.
Feoli$^{d,e,f}$, G. Lambiase$^{d,e}$}
 \address{$^a$Department of Physics, University of Regina, Regina, Sask, S4S 0A2, Canada}
 \address{$^b$Prairie Particle Physics Institute, Regina, Sask,
S4S 0A2, Canada}
  \address{$^c$International Institute for Advanced Scientific Studies, 89019 Vietri sul Mare (SA), Italy}
 \address{$^d$Dipartimento di Fisica "E.R. Caianiello"
 Universit\'a di Salerno, 84081 Baronissi (SA), Italy}
  \address{$^e$INFN - Gruppo Collegato di Salerno, Italy}
\address{$^f$Dipartimento d'Ingegeneria, Universit\'a del Sannio,
Corso Garibaldi 107, Palazzo Bosco Lucarelli, 82100 Benevento,
Italy.}
\ead{papini@uregina.ca,scarpetta@sa.infn.it,feoli@unisannio.it,\\
lambiase@sa.infn.it}

\begin{abstract}
The Maxwell and Maxwell-de Rham equations can be solved exactly to
first order in an external gravitational field. The gravitational
background induces phases in the wave functions of spin-$1$
particles. These phases yield the optics of the particles without
requiring any thin lens approximation.
\end{abstract}


\maketitle

\section{Introduction}

The deflection of photons from distant astronomical sources by
intervening massive objects is rapidly becoming an important tool
for experimental astrophysics and cosmology. At the same time a
variety of optical, atomic and molecular interferometers in use or
under development is approaching the threshold of measurability
for inertial and gravitational effects.

Photons in gravitational fields are described by the generalized
Maxwell equations
\begin{equation}\label{1}
  \nabla_\alpha \nabla^\alpha A_\mu = 0\,,
\end{equation}
or, more generally, by the Maxwell-de Rahm equations
\begin{equation}\label{2}
  \nabla_\alpha \nabla^\alpha A_\mu - R_{\mu\alpha} A^\alpha = 0\,,
\end{equation}
where $\nabla_\alpha$ indicates covariant differentiation. We use
units $ \hbar=c=1$.

In interferometry, where the phase shifts are proportional to the
mass of the particle employed, one may wish to consider, beside
photons, also massive, charge-less, spin-1 particles that satisfy
the generalized Proca equation
\begin{equation}\label{3}
  \nabla_\alpha \nabla^\alpha A_\mu +m^2 A_\mu = 0\,,
\end{equation}
in the hope that among the atoms and molecules to be used some
indeed obey (\ref{3}).

We have shown in previous work \cite{cai1}-\cite{punzi}, that
equations (\ref{1})-(\ref{3}) can be solved exactly to first order
in the metric deviation $\gamma_{\mu\nu}=g_{\mu\nu}-\eta_{\mu\nu}$,
where $|\gamma_{\mu\nu}|\ll 1$ and $\eta_{\mu\nu}$ is the Minkowski
metric of signature -2. In all these solutions, the background
gravitational field appears in a phase operator that acts on the
wave function of the relative field-free equation. It is precisely
from the corresponding gravity-induced phases that the optics of the
particles considered follows.

The purpose of this paper is to show that the use of the solutions
found presents advantages, in principle, over other current
approximations. This is particularly true of gravitational lensing
where large use is made of the thin lens approximation
\cite{schn},\cite{suy} in which the lens is characterized by a
surface mass density and the waves are scattered only at the thin
lens plane. The phases can in fact be calculated by means of path
integrals that can be integrated easily and exactly for most known
metrics. Gravity acts along the complete length of the space-time
paths and not only in a narrow region at the lens plane
\cite{pap2}.

The plan of the paper is as follows. In Section II we give, for
completeness, the solutions of (\ref{1})-(\ref{3}) and show, in
Section III, that they are gauge invariant. Geometrical and wave
optics are discussed in Section IV. We summarize the results in
Section V.

\section{\label{sec:2}Solution of the spin-1 wave equation}

To first order in $\gamma_{\mu\nu}$, (\ref{1})-(\ref{3}) become

\begin{equation}\label{4}
  \nabla_\nu \nabla^\nu A_\mu \simeq
  (\eta^{\sigma\alpha}-\gamma^{\sigma\alpha})A_{\mu,\alpha\sigma}-
  2 \Gamma_{\sigma,\,\mu\nu}
  A^{\sigma,\nu}-\frac{1}{2}\gamma_{\sigma\mu,\nu}^{\,\,\,\,\,\,\,\,\nu}A^\sigma =
  0\,,
 \end{equation}
  \begin{equation}\label{5}
  \nabla_\nu\nabla^\nu A_\mu -R_{\mu\alpha}A^\alpha \simeq
  (\eta^{\sigma\alpha}-\gamma^{\sigma\alpha})A_{\mu,\alpha\sigma}+
  2 \Gamma_{\sigma,\,\mu\nu}
  A^{\sigma,\nu} = 0\,,
  \end{equation}
 \begin{equation}\label{6}
  (\nabla_\nu \nabla^\nu+m^2)A_\mu \simeq
  (\eta^{\sigma\alpha}-\gamma^{\sigma\alpha})A_{\mu,\alpha\sigma}+
  2 \Gamma_{\sigma,\,\mu\nu}
  A^{\sigma,\nu}-
      \frac{1}{2}\gamma_{\sigma\mu,\nu}^{\,\,\,\,\,\,\,\,\nu}A^\sigma +m^2 A_\mu =
  0\,,
  \end{equation}
where ordinary differentiation of a quantity $\Phi$ is equivalently
indicated by $\Phi_{,\alpha}$ or $\partial_\alpha \Phi$ and $
\Gamma_{\sigma,\,\mu\nu}=1/2(\gamma_{\sigma\mu,\nu}+\gamma_{\sigma\nu,\mu}-\gamma_{\mu\nu,\sigma})$
are the Christoffel symbols. In what follows we also use the
notations $ K^{\alpha}\equiv x^{\alpha}-z^{\alpha}$ and $
B_{[\alpha\lambda...\beta]}\equiv
B_{\alpha\lambda...\beta}-B_{\beta\lambda...\alpha}$. In deriving
(\ref{4})-(\ref{6}), we have used the Lanczos-De Donder gauge
condition
\begin{equation}\label{7}
  \gamma_{\alpha\nu,}^{\,\,\,\,\,\,\,
  \nu}-\frac{1}{2}\gamma_{\sigma,\alpha}^\sigma = 0\,.
\end{equation}
The field $A_\mu(x)$ also satisfies the condition
\begin{equation}\label{8}
  \nabla_\mu A^\mu=0\,.
\end{equation}
Though (\ref{8}) is strictly required only when $m=0$, it is
convenient to impose it also in the case of a massive spin-1
particle. Equations (\ref{4}) and (\ref{6}) can be handled
simultaneously. As shown in Appendix A, their solution is
 \begin{equation}\label{9}
A_\mu(x)  \simeq  e^{-i\xi}a_\mu(x)\approx (1-i\xi)a_\mu(x) =
a_{\mu}(x)+\frac{1}{2}\int_{P}^{x}dz^{\lambda}
\gamma_{\alpha\lambda}(z)
 \partial^{\alpha}a_{\mu}(x)
\end{equation}
 \[
 -\frac{1}{4}\int_{P}^{x}dz^{\lambda} \gamma_{[\alpha\lambda,\beta]}(z)
 K^{[\alpha}\partial^{\beta]}a_{\mu}(x)
 +\int_{P}^{x}dz^{\lambda}\Gamma_{\sigma,\,\mu\lambda}(z)
 a^{\sigma}(x)\,,
 \]
where $a_\mu$ satisfies the equation $\partial_\nu\partial^\nu a_\mu
=0$ in the case of (\ref{4}), and $(\partial_\nu\partial^\nu +
m^2)a_\mu =0$ when (\ref{9}) is a solution of (\ref{6}). The
solutions $ a_{\mu}(x)$ used in this paper (Section IV) are either
plane waves or spherical waves multiplied by $ e^{-ik_{0}x^{0}}$.
For this reason, it is convenient to group $ i$ with $ \xi $.

Similarly, it can be verified by substitution that the Maxwell-de
Rahm equation (\ref{5}) has the solution \cite{pap1}
\begin{equation}\label{10}
A_\mu(x) \simeq  e^{-i\eta}a_\mu(x)\approx (1-i\eta)a_\mu(x)
=a_\mu  +\int_P^x dz^\lambda \gamma_{\alpha\mu}(z)
\partial_\lambda a^\alpha (x)-
 \end{equation}
 \[
  - \frac{1}{4}\int_P^x dz^\lambda\left[
 \gamma_{[\alpha\lambda,\beta]}(z) K^{[\alpha}\partial^{\beta]}\eta_{\beta\mu}-
 2 \gamma_{\alpha \lambda}\partial^\alpha \eta_{\beta\mu}+
 2\gamma_{[\mu\lambda,\beta]}(z)\right]a^\beta(x)\,,
 \]
where again $\partial_\alpha \partial^\alpha a_\mu =0$ and
\begin{equation}\label{11}
 -i\eta a_\mu (x)=-i\xi a_\mu(x)
-\frac{\epsilon}{2}\int_P^x dz^\lambda[
 \gamma_{\alpha\mu,\lambda}(z) a^{\alpha}(x)-2 \gamma_{\alpha \mu}(z)
\partial_\lambda] a^\alpha (x)\,.
\end{equation}
The last two terms are typical of the Maxwell-de Rahm solution and
have been tagged with the parameter $\epsilon$ ($\epsilon =1$) for
bookkeeping purposes. Solutions (\ref{9}) and (\ref{10}) must also
satisfy (\ref{8}). On using $\nabla^{\alpha}A_{\alpha}\simeq
\eta^{\alpha\beta}A_{\alpha,\beta}-\gamma^{\alpha\beta}a_{\alpha,\beta}$
and (\ref{A.3}), we obtain
\begin{equation}\label{11bis}
  \nabla^{\alpha}A_\alpha= f(x)
  \end{equation}
 \[
 -f(x) \equiv   \frac{1}{2}\gamma_{\sigma\mu}(a^{\sigma,\mu}+a^{\mu,\sigma})
  -\frac{1}{2}\partial^\mu(\gamma_{\sigma\mu} a^\sigma)+\epsilon
  \left[\frac{dx^\lambda}{ds}\gamma_{\alpha\mu} (x)
  \partial^\mu\partial_\lambda a^\lambda(x)\right]_{s_1}^{s_2} \,.
\]
By performing the gauge transformation $A_\sigma\to A_\sigma +
\Lambda_{, \sigma}$, we find $\nabla^\sigma A_\sigma\to
\nabla^\sigma A_\sigma+\nabla^\sigma \nabla_\sigma \Lambda + f=0$,
from which we obtain $\nabla^\sigma A_\sigma =0$ provided
\begin{equation}\label{12}
 \nabla^\sigma \nabla_\sigma \Lambda + f=0\,.
\end{equation}

Finally, the addition of a term $m^2 A_\mu$ to the l.h.s. of
(\ref{2}) invalidates the solution (\ref{10}) that can not,
therefore, be extended to include de Rahm's term, except when $ 2
\epsilon m^2\gamma_{\mu\nu}a^{\nu}$ can be considered a second
order quantity. It is known, on the other hand, that the term $
R_{\mu\alpha}A^{\alpha}$ can be dropped in lensing when the
wavelength $ \lambda$ of $ A^{\alpha}$ is much smaller than the
typical radius of curvature of the gravitational background, or
when, in the weak field approximation, $ R_{\mu\alpha}= - 1/2
\gamma_{\mu\alpha,\nu}^{\,\,\,\,\,\,\,\,\nu}=0 $.

\section{\label{sec:3}Gauge invariance}

The first two integrals in (\ref{9}) and (\ref{10}) represent by
themselves a solution of the Klein-Gordon equation $(\nabla_\mu
\nabla^\nu + m^2)\phi =0$. The additional terms must therefore be
related to spin. In fact (\ref{9}) can be written as
\begin{equation}\label{14}
  A_\mu = a_\mu -\frac{1}{2}\int_P^x dz^\lambda
  \gamma_{[\alpha\lambda,\beta]}K^\alpha\partial^\beta
  a_\mu +\frac{1}{2}\int_P^x dz^\lambda
  \gamma_{\alpha\lambda}\partial^\alpha a_\mu+
   \end{equation}
 \[
  +\frac{i}{2}\int_P^x
  dz^\lambda
  \gamma_{[\alpha\lambda,\beta]}S^{\alpha\beta}a_\mu
  +\frac{i}{2}\int_P^x dz^\lambda
  \gamma_{\alpha\beta,\lambda}T^{\alpha\beta}a_\mu (x)\,,
 \]
where the tensors
\begin{equation}\label{15}
  (S^{\alpha\beta})_{\mu\nu}=-\frac{i}{2}(\delta^\alpha_\mu
  \delta_\nu^\beta -\delta_\mu^\beta \delta_\nu^\alpha)\,,
  \quad
 (T^{\alpha\beta})_{\mu\nu}=-\frac{i}{2}(\delta^\alpha_\mu
  \delta_\nu^\beta +\delta_\mu^\beta \delta_\nu^\alpha)\,
\end{equation}
have been written in matrix form as $S^{\alpha\beta}$ and
$T^{\alpha\beta}$ and $S^{\alpha\beta}a_\mu=(S^{\alpha\beta})_{
\mu\nu} a^\nu$, where $a^{\nu}$ is a column matrix. The matrices
$S_i=2\epsilon_{ijk}S^{jk}$ satisfying the commutation relation
$[S_i, S_j]=i\epsilon_{ijk}S_k$, can be recognized as rotation
matrices. These, and the matrix $ a^\nu$, are represented by
 \[
 S_1=-i\left(\begin{array}{cccc}
 0 & 0 & 0 & 0 \vspace{0.05in} \\
 0 & 0 & 0 & 0 \vspace{0.05in} \\
 0 & 0 & 0 & 1 \vspace{0.05in} \\
 0 & 0 & -1 & 0 \vspace{0.05in} \\
\end{array}\right)\,,
 S_2=i\left(\begin{array}{cccc}
 0 & 0 & 0 & 0 \vspace{0.05in} \\
 0 & 0 & 0 & 1 \vspace{0.05in} \\
 0 & 0 & 0 & 0 \vspace{0.05in} \\
 0 & -1 & 0 & 0 \vspace{0.05in} \\
\end{array}\right)\,,
 \]
 \[
 S_3=-i\left(\begin{array}{cccc}
 0 & 0 & 0 & 0 \vspace{0.05in} \\
 0 & 0 & 1 & 0 \vspace{0.05in} \\
 0 & -1 & 0 & 0 \vspace{0.05in} \\
 0 & 0 & 0 & 0 \vspace{0.05in} \\
\end{array}\right)\,,
a^\nu=\left(\begin{array}{c}
 a^0  \vspace{0.05in} \\
 a^1 \vspace{0.05in} \\
 a^2 \vspace{0.05in} \\
 a^3 \vspace{0.05in} \\
 \end{array}\right)\,.
 \]
We give the matrices $ S^{\alpha\beta}$ in Appendix B and show the
equivalence of matrix and tensor expressions. In the same Appendix
we also show that by applying Stokes theorem to the line integrals
on the r.h.s. of (\ref{14}) extended to a closed space-time path
$\Gamma$, we obtain, to first order \cite{landau},
\begin{equation}\label{16}
  A_\mu = \left(1-\frac{i}{4}\int_\Sigma d\tau^{\sigma\delta}R_{\sigma\delta \alpha
  \beta}J^{\alpha\beta}\right) a_\mu \,,
\end{equation}
where $J^{\alpha\beta}=L^{\alpha\beta}+S^{\alpha\beta}$ is the total
angular momentum of the spin-1 particle, $R_{\mu\nu\alpha\beta}=1/2
(\gamma_{\mu\beta,\nu\alpha}+\gamma_{\nu\alpha,\mu\beta}-\gamma_{\mu\alpha,\nu\beta}-
\gamma_{\nu\beta,\mu\alpha})$ is the linearized Riemann tensor, and
$\Sigma$ is the surface bound by $\Gamma$. Equation (\ref{16}) is
unambiguous because, to first order, the values of $
\gamma_{\alpha\beta}$ on $ \Sigma$ can be obtained from those on
$\Gamma$ independently of the path followed \cite{landau}. The term
of (\ref{14}) that contains $ S_{\alpha\beta} $ gives rise to the
Skrotskii effect \cite{skr},\cite{kop}. It follows from (\ref{16})
that (\ref{9}) is covariant and also invariant under the gauge
transformations $\gamma_{\mu\nu}\to
\gamma_{\mu\nu}-\xi_{\mu,\nu}-\xi_{\nu,\mu}$ induced by the
coordinate transformations $x_\mu\to x_\mu+\xi_\mu$ which are still
allowed in the weak field approximation. The quantities $\xi_\mu$,
also small of first order, must satisfy the equation $
\xi_{\mu,\sigma}^{\,\,\,\,\,\sigma}=0 $ because of the Lanczos-De
Donder condition. It also follows from (\ref{16}) that the term
containing $T^{\alpha\beta}$ in (\ref{14}) does not contribute to
integrations over closed paths, behaves as a gauge term and may
therefore be dropped.

The difference between (\ref{10}) and (\ref{9}) is represented by
the two terms labelled by $ \epsilon $ in (\ref{11}). Of these,
the first one vanishes when integrated over a closed path. To
evaluate the second term we apply the transformations
$\gamma_{\mu\nu}\to \gamma_{\mu\nu}-\xi_{\mu,\nu}-\xi_{\nu,\mu}$,
integrate over a closed path and choose, for simplicity,
$\xi_\alpha = \xi_{\alpha 0}e^{-ilx}$ and $a_\alpha = a_{\alpha
0}e^{-ikx}$, where $l_\alpha l^\alpha =0$, $k_\alpha k^\alpha=0$.
We obtain
 \[
 -i\epsilon \oint dz^\lambda k_{\lambda}(\xi_{\alpha,\mu} (z)+\xi_{\mu,
 \alpha}(z))a^\alpha (x)=
 -\frac{\epsilon}{2}l_{[\mu} \xi_{\alpha]\, 0}a^\alpha(x) \oint dz^\lambda l_\lambda +
 \]
 \[
 +i\epsilon \left[l_\mu \oint dz^\lambda (\partial_\lambda a^\alpha
 (x) \xi_{\alpha 0})+\xi_{\mu 0} \oint dz^\lambda (\partial_\lambda a^\alpha
 (x) l_\alpha)\right]=0\,,
 \]
because both integrands are of the type $du$ and return to the
initial value at the final point. This shows that the addition of
the de Rahm term to the solution does not affect its invariance
under the gauge transformations of $ \gamma_{\mu\nu}$.

We must now consider the effect of the electromagnetic gauge
transformations $a_\mu\to a_\mu + \Lambda_{,\mu}$ which have already
been used to obtain (\ref{12}). Equation (\ref{9}) can be explicitly
re-written in term of the gauge invariant electromagnetic field
tensors $F_{\mu\nu}=\nabla_\mu A_\nu - \nabla_\nu A_\mu = A_{[\nu,
\mu]}$, and $f_{\mu\nu}=a_{[\nu, \mu]}$ as follows
 \[
 F_{\mu\nu}= f_{\mu\nu} -\frac{1}{2}\int_P^x dz^\lambda
 \gamma_{[\alpha \lambda,\beta]}(z)
 k^\alpha \partial^\beta f_{\mu\nu}(x)+
  \int_P^x dz^\lambda \gamma_{[\mu\lambda,}^{\,\,\,\,\,\,\,\,\beta]}(z)f_{\nu\beta}(x)
 \]
 \[
 +\frac{1}{2}\int_P^x
 dz^\lambda \gamma_{\alpha\lambda} (z) \partial^\alpha
 f_{\mu\nu}(x)+\gamma^\beta_\nu (x) f_{\mu\beta}(x)\,.
 \]
The additional terms that the Maxwell-de Rham solution (\ref{10})
has relative to (\ref{9}) can be re-written in the form
\begin{equation}\label{18}
  -\frac{\epsilon}{2}\int_P^x dz^\lambda \left[
  \gamma_{\beta\mu, \lambda}(z) a^\beta (x)-
  2 \gamma^\alpha_\mu (z) f_{\lambda \alpha}(x)-
  2 \gamma^\alpha_\mu (z) a_{\lambda, \alpha}(x)\right]\,.
 \end{equation}
The second term of (\ref{18}) is already invariant because it
contains $ f_{\lambda\alpha}$. The first term vanishes when the
integration is over closed space-time paths. The last term
transforms according to
\begin{equation}\label{19}
  \epsilon \int_P^x dz^\lambda \gamma_\mu^\alpha (z) a_{\lambda,
  \alpha}(x)\to \epsilon \int_P^x dz^\lambda \gamma_\mu^\alpha (z) (a_{\lambda,
  \alpha}+\Lambda_{, \lambda \alpha})\,,
 \end{equation}
can be re-written in the form
 \[
 \epsilon \int_P^x dz^\lambda \frac{\partial^2}{\partial x^\lambda \partial x^\alpha}
 (\gamma_{\alpha \mu}(z) \Lambda(x))\,,
 \]
and vanishes when the integration is over closed space-time paths.
Equation (\ref{10}) is therefore invariant under electromagnetic
gauge transformations.

\section{\label{sec:4}Optics}

\subsection{\label{sec:4A}Lensing}

The correct general relativistic deviation of light rays in the
gravitational field of a star follows immediately from the first
two integrals in (\ref{9}) and (\ref{10}). We choose, e.g.,  a
gravitational background represented by the Lense-Thirring metric,
$\gamma_{00}=2\phi$, $\gamma_{ij}=2\phi\delta_{ij}$, $\phi=-GM/r$,
and $\gamma_{0i}\equiv h_i = 2GJ_{ij}x^j/r^3$, where $x^j=(x, y,
z)$, $r=\sqrt{x^2+y^2+z^2}$ and $J_{ij}$ is related to the angular
momentum of the gravitational source. In particular, if the source
rotates with angular velocity ${\vec \omega}=(0,0,\omega)$, then
$h_1=4GMR^2\omega y/5r^3$, $h_2=-4GMR^2\omega x/5r^3$, .

Without loss of generality, we assume that photons propagate along
the $z$-direction, hence $k^\alpha\simeq (k, 0, 0, k)$. Moreover,
photons propagate along null geodesics, from which it follows that
$ds^2=0$, or $dt=dz$. Using plane waves for
$a_\mu(x)=a_{\mu}^{0}exp(-i k_{\alpha}x^{\alpha})$, the solution has
the form $A_\mu=e^{-i\chi}a_\mu^0$, where
\begin{equation}\label{gph}
 \chi =k_{\alpha}x^{\alpha} -\frac{1}{4}\int_P^x dz^\lambda
 \gamma_{[\alpha\lambda,\beta]}(z)K^{[\alpha}k^{\beta]}+
 \frac{1}{2}\int_P^x dz^\lambda \gamma_{\alpha\lambda}(z)
 k^\alpha\,.
 \end{equation}
We can define the photon momentum as
\begin{equation}\label{photonmomentum}
 \tilde{k}_\alpha = \frac{\partial \chi}{\partial x^\alpha}\,.
\end{equation}
It is easy to show that $ \chi $ satisfies the eikonal equation $
g^{\alpha\beta} \chi_{,\alpha} \chi_{,\beta}=0 $. We can also
prove, by direct substitution, that the inclusion of de Rahm's
terms (see (\ref{11})) in (\ref{gph}) does not invalidate the
eikonal equation. Geometrical optics is not therefore affected by
the term $ R_{\mu\alpha}A^{\alpha}$ in (\ref{2}).

For the Lense-Thirring metric, $\chi$ is given by
\begin{eqnarray}\label{chiLT}
 \chi & \simeq & -\frac{k}{2}\int_P^Q \{
   (x-x')\phi_{, z'}dx'+(y-y')\phi_{,\, z'}dy'- \\
 & & - 2[(x-x')\phi_{,\, x'}+(y-y')\phi_{,
   \, y'}]dz'\} +  k\int_P^Q dz' \phi -\nonumber \\
   & - & \frac{k}{2}\int_P^Q \left\{\left[
    (x-x')(h_{1, \,z'}-2h_{3,\, x'})+(y-y')(h_{2, \,z'}-2h_{3,\, y'})\right] dz' -\right. \nonumber \\
    & - & \left. \left[(x-x')h_{1,\, x'}+(y-y')h_{1,\, y'}\right]dx'
   -\left[(x-x')h_{2,\, x'}+(y-y')h_{2,\, y'}\right]dy'\right\} \nonumber \\
  & + & \frac{k}{2}\int_P^Q\left[2h_3dz'+h_1dx'+h_2dy'\right]\,,
\end{eqnarray}
where $P$ is the point at which the photons are generated, and $Q$
is a generic point along their trajectory.  The components of the
photon momentum are therefore
\begin{eqnarray}\label{k1}
 \tilde{k}_1 &=& 2k\int_P^Q  \left(-\frac{1}{2}\frac{\partial \phi}{\partial z}\,dx -
 \frac{1}{2}\frac{\partial h_2}{\partial x}\,dy +\frac{\partial (\phi+h_3)}{\partial
 x}dz\right)- \nonumber \\
 & &  -\frac{k}{2}(h_1(Q)-h_1(P))\,,\\
 \tilde{k}_2 &=& 2k\int_P^Q  \left(-\frac{1}{2}\frac{\partial \phi}{\partial z}\,dy
 +\frac{1}{2}\frac{\partial h_1}{\partial y}\,dx +
 \frac{\partial (\phi+h_3)}{\partial y}dz\right)+ \nonumber \\
 & & +\frac{k}{2}(h_2(Q)-h_2(P))\,, \label{k2} \\
  \tilde{k}_3 &=& k(1+\phi+h_3)\,. \label{k3}
\end{eqnarray}
We also have
  ${\bf \tilde{k}}={\bf \tilde{k}}_\perp +\tilde{k}_3\, {\bf e}_3\,, \quad {\bf
  \tilde{k}}_\perp=\tilde{k}_1\, {\bf e}_1+\tilde{k}_1\, {\bf e}_2\,$,
where ${\bf \tilde{k}}_\perp$ is the component of the momentum
orthogonal to the direction of propagation of the photons.

Since only phase differences are physical, it is convenient to
choose the space-time path by placing the photon source at
distances that are very large relative to the dimensions of the
lens, and the generic point is located along the $z$ direction. We
therefore replace $Q$ with $z$, where $z\gg x, y$. Using the
expression for $h_{1,2}$ we find that their contribution is
negligible and (\ref{k1})-(\ref{k3}) simplify to
\begin{eqnarray}\label{k1bis}
 \tilde{k}_1 &=& 2k\int_{-\infty}^z  \frac{\partial (\phi+h_3)}{\partial x}dz\,,\\
 \tilde{k}_2 &=& 2k\int_{-\infty}^z  \frac{\partial (\phi+h_3)}{\partial y}dz\,, \label{k2bis} \\
  \tilde{k}_3 &=& k(1+\phi+h_3)\,. \label{k3bis}
\end{eqnarray}
From (\ref{k1bis})-(\ref{k3bis}) we can determine the deflection
angle $\theta$. Let us analyze first the case of non-rotating
lenses, i.e. $h_3=0$. We get
\begin{eqnarray}\label{k1tris}
 \tilde{k}_1 &\sim & k\, \frac{2GM}{R^2}\, x\left(1+\frac{z}{r}\right)\,, \\
 \tilde{k}_2 &\sim &  k\, \frac{2GM}{R^2}\, y\left(1+\frac{z}{r}\right) \,,\label{k2tris} \\
  \tilde{k}_3 &=& k(1+\phi+h_3)\,, \label{k3tris}
\end{eqnarray}
where $R=\sqrt{x^2+y^2}$. By defining the deflection angle as
  $\tan \theta = \tilde{k}_\perp/\tilde{k}_3\,$,
it follows that
\begin{equation}\label{thetak}
  \tan \theta \sim \theta \sim
  \frac{2GM}{R}\left(1+\frac{z}{r}\right)\,.
\end{equation}
In the limit $z\to \infty$ we obtain the usual Einstein result
 $ \theta_M\sim 4GM/R $.
The remaining term of (\ref{14}) contains $S_{\alpha\beta}$ and
therefore is a spin effect. It is usually referred to as the
Skrotskii effect and represents a rotation of the plane of
polarization of the incoming photon beam.

A general expression for the index of refraction $ n $ can also be
derived from (\ref{gph}),(\ref{photonmomentum}) and  $ n =
\tilde{k}/\tilde{k}_0 =\chi_{,3}/\chi_{,0}$.

\subsection{\label{sec:4B}Wave effects in gravitational lensing}

We consider the propagation of light waves in a background metric
represented by $\gamma_{00}=2U(\rho)$, $\gamma_{ij}=2 U(\rho)
\delta_{ij}$, where $U(\rho)=-GM/\rho$ is the gravitational
potential of the lens. Wave optics effects can be calculated by
using the type of double slit arrangement indicated in Fig.1. We
neglect all spin effects. This limits the calculation of the phase
difference to the first two terms in (\ref{9}) and (\ref{10}). We
also assume for simplicity that $k^1 =0$, so that propagation is
entirely in the $(x^2, x^3$)-plane and the set-up is planar. We use
a solution of the equation $\partial^\gamma
\partial_\gamma a_\alpha =0$ of the form
$a_\mu = a_\mu^0e^{-ik_{\sigma}x^{\sigma}}/r=a_\mu^0
e^{-ik_{0}x^{0}}e^{-ik_i x^i}/r=a_\mu^0 e^{-ik_{0}x^{0}}e^{ikr}/r$,
valid everywhere outside the light source ($ r=0$), which is the
region of physical interest in lensing. We neglect the contribution
from the de Rahm term because, for the metric used,
$R_{\sigma\mu}A^{\sigma}=-1/2
\gamma_{\sigma\mu,\nu}^{\,\,\,\,\,\,\,\,\nu}A^{\sigma}=
-\nabla^{2}(GM/\rho)A_{\mu}$ which vanishes for $ \rho\neq 0$.

The corresponding wave amplitude $\phi$ is therefore
\begin{equation}\label{*}
 \phi(x) = \frac{e^{-i k_\sigma x^\sigma}}{2r}\left\{2+
 \int_S^x \left[ dz^0 \gamma_{00}
 \Pi^0+ dz^2 \gamma_{22} \Pi^2+ dz^{3}\gamma_{33}\Pi^{3}\right]- \right.
 \end{equation}
 \[
 \left. \int_S^x \left[dz^0 \left(\gamma_{00,2} K^{[0} \Pi^{2]} +
  \gamma_{00, 3}K^{[0} \Pi^{3]}\right)+
 dz^2 \gamma_{22,3}K^{[2}\Pi^{3]}+
 dz^3 \gamma_{33,2}K^{[3}\Pi^{2]}\right]\right\}
 \]
where $\Pi^0=-i k$, $\Pi^i = -ik^i -x^i/r$, and we have taken into
account the fact that $\gamma_{11}$ plays no role in the planar
arrangement chosen. The change in phase must now be calculated along
the different paths SP+PO and SL+LO according to $\Delta
\tilde{\phi} =
(\phi_{SLO}-\phi_{SPO})/(e^{-ik_{\sigma}x^{\sigma}}/r)$ and taking
into account the values of $\Pi^i$ in the various intervals.
\begin{figure}
\centering
\includegraphics[width=0.9\textwidth]{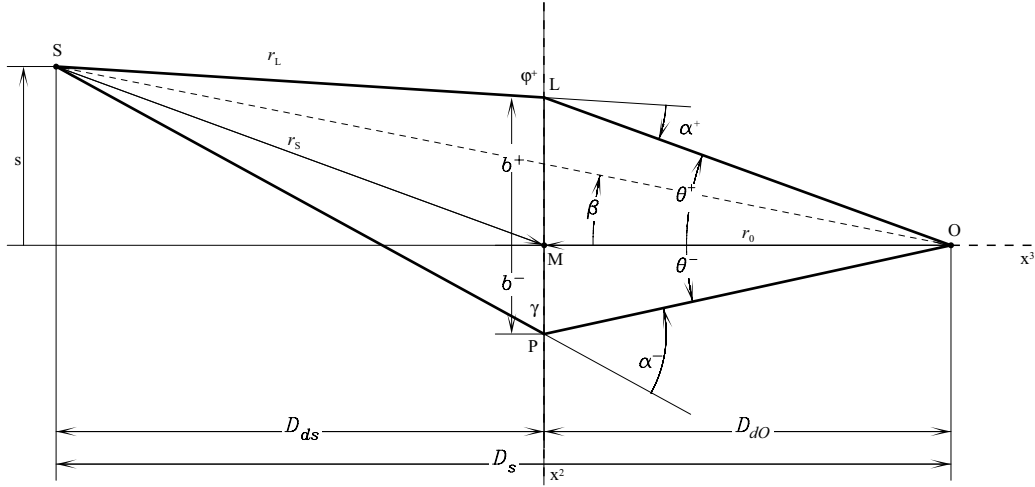}
\caption{\label{fig:Lensing} Geometry of a two-image gravitational
lens or, equivalently, of a double slit interference experiment.
The solid lines represent the particle paths between the particle
source at $S$ and the observer at $O$. $M$ is the spherically
symmetric gravitational lens. $S, M, O$ and the particle paths lie
in the same plane. The physical variables are $ r_{S}, r_{0},
b^{\pm}, s$, while the lensing variables are indicated by $
D_{dS}, D_{dO}, D_{S}, \theta^{\pm}, \beta$. The deflection angles
are indicated by $ \alpha^{\pm}$.}
\end{figure}
The total change in phase is
 $ \Delta \tilde{\phi}= \Delta \tilde{\phi}_{SL} +\Delta \tilde{\phi}_{LO}-\Delta \tilde{\phi}_{SP}-
 \Delta \tilde{\phi}_{PO}$.
 The details of the calculation of $\Delta\tilde{\phi}$ along the
path segments are given in Appendix C. All integrations in
(\ref{3*}), (\ref{5*}). (\ref{7*}) and (\ref{9*}) can be performed
exactly.
 All results can be expressed in terms of physical variables
 $r_s$, $r_0$, $b^+$, $b^-$, and $s$ and lensing variables $D_s$,
 $D_{ds}$, $D_d$, $\theta^+$, $\theta^-$, and $\beta$. We obtain
 \begin{eqnarray}\label{FR}
\Delta\tilde{\phi} & = & \tilde{y} \left\{\ln
\left(-\sqrt{D_{dS}^2 +\left(s + b^-\right)^2} + b^- \cos\gamma +
r_S \right)- \ln \left(b^- \left(1+ \cos\gamma\right)\right)
\right.\nonumber \\
& + & \ln \left(b^+ \left(1-\cos\varphi^+ \right)\right)-
\ln\left(r_S -r_L-b^+\cos\varphi^+\right)
\nonumber \\
& + & \ln \left( b^- + r_0 \cos\theta^- -\sqrt{b^{-\,2}+
r_0^2}\right)-\ln \left(r_0\left(1+ \cos\theta^-\right)\right)
\nonumber \\
& + & \left.\ln\left(r_0\left(1+ \cos\theta^+\right)\right)-
\ln\left(b^+ +r_0 \cos\theta^+ -\sqrt{b^{+\,2}+r_0^2}\right)
\right\} \,,
\end{eqnarray}
where $ r_S^2 = b^{+\,2}+r_L^2+2b^+r_L\cos\varphi^+$\,,
$r_L^2=D_{dS}^2 +(s-b^+)^2$\,, $ \varphi^+ +\alpha^+ +\alpha^-
+\gamma -\theta^+ -\theta^- =\pi$ and $ \tilde{y}= 2GMk$. Equation
(\ref{FR}) agrees with that obtained in \cite{pap2} for plane
waves and can be applied to the optics of any particle when spin
effects are neglected.

The contributions $ \tilde{\chi} $ of the real parts of $\Pi^{2}$
and $ \Pi^{3}$ to (\ref{3*}), (\ref{5*}), (\ref{7*}) and
(\ref{9*}) are listed below by path segment
\begin{eqnarray} \label{A1}
\tilde{\chi}_{SL}&=&\frac{GM}{r_L^2}\left\{b^+ \frac{-2+ 3
\cos^2\varphi^{+}}{\cos\varphi^+} \ln\frac{r_S-r_L-b^+
\cos\varphi^+}{b^+ \left(1-\cos\varphi^+\right)}-\right. \\
& &  \left. -\frac{2r_Lb^+ \sin^2\varphi^{+}}{r_S
\cos\varphi^+}\right\}\,, \nonumber
 \end{eqnarray}
\begin{eqnarray}\label{A2}
\tilde{\chi}_{LO}&=&\frac{GM}{R_{1}^2}\left\{r_0 \frac{2-
3\cos^2\theta^{+}}{\cos\theta^+} \ln\frac{b^+ -\sqrt{r_0^2
+b^{+\,2}}+b^+\cos\theta^+}{r_0 \left(1+\cos\theta^+\right)}+ \right. \\
 && \left.
+ \frac{2r_0\sqrt{r_0^2 +b^{+\,2}} \sin^2\theta^{+}}{b^+
\cos\theta^+}\right\}\,, \nonumber
\end{eqnarray}

\begin{eqnarray}\label{A3}
\tilde{\chi}_{SP}&=&\frac{GM}{R^2}\left\{b^- \frac{2-
3\cos^2\gamma}{\cos\gamma} \ln\frac{r_S -\sqrt{D_{dS}^2
+\left(s+b^-\right)^2}+b^-\cos\gamma}{b^- \left(1+\cos\gamma\right)}+ \right. \\
 & & \left. +\frac{2b^-\sqrt{D_{dS}^2 +\left(s+b^-\right)^2} \sin^2\gamma}{r_S
\cos\gamma}\right\}\,, \nonumber
\end{eqnarray}

\begin{eqnarray}\label{A4}
\tilde{\chi}_{PO}&=&\frac{GM}{R_{2}^2}\left\{r_0 \frac{2-
3\cos^2\theta^{-}}{\cos\theta^-} \ln\frac{b^- -\sqrt{r_0^2
+b^{-\,2}}+r_0\cos\theta^-}{r_0 \left(1+\cos\theta^-\right)}+
\right.\\
 & & \left. + \frac{2r_0\sqrt{r_0^2 +b^{-\,2}} \sin^2\theta^{-}}{b^-
\cos\theta^-}\right\}\,. \nonumber
\end{eqnarray}
Contrary to $ \Delta\tilde{\phi}$, $
\tilde{\chi}=\tilde{\chi}_{SL}+\tilde{\chi}_{LO}+\tilde{\chi}_{SP}+\tilde{\chi}_{PO}$
is completely independent of $ \tilde{y}$.

The meaning of (\ref{A1})-(\ref{A4}) becomes evident by recalling
that the total wave amplitude has the form
\begin{equation}\label{wf}
\phi(x)= \frac{e^{ikr}}{r}
\exp\left[-ik_{0}x^{0}-i\xi\left(x\right)+
\tilde{\chi}\left(x\right)\right]\,
\end{equation}
and that the ratio
\begin{equation}\label{mf}
F\equiv \frac{\phi
\phi^{\ast}}{\phi_{0}\phi_{0}^{\ast}}=e^{\tilde{\chi}+\tilde{\chi}^\ast}
\end{equation}
is known as the amplification factor. As an example, let us
consider the simpler case of particles coming from $ x^{3}
=-\infty\,\, (\varphi^+=\gamma=\pi/2)$ with $ b^{+}= b^{-}=s\equiv
b\,,r_0=D_{dS}$. Then Fig.1 gives $
\theta^{+}=\theta^{-}\equiv\theta\,,$ and $R_1=R_2$ . In this case
$
\tilde{\chi}_{SL}=-\tilde{\chi}_{SP}\,,\tilde{\chi}_{LO}=\tilde{\chi}_{PO}$
and $ \Delta\tilde{\phi}=0$. A simple calculation shows that the
probability density of finding a particle at $ O$ is
\begin{equation}\label{PD}
\phi\phi^* \propto \frac{4}{r^2}
e^{\tilde{\chi}+\tilde{\chi}^\ast}\cos^2
\frac{\Delta\tilde{\phi}}{2}=\frac{4}{r^2}e^{\tilde{\chi}+\tilde{\chi}^\ast}\,,
\end{equation}
where
\begin{equation}\label{chit}
\tilde{\chi}=\frac{4GM}{R_1}\left\{\sin\theta+\frac{2-3\cos^2\theta}{2}\ln\frac{\sin\theta-1+
\sin\theta\cos\theta}{\cos\theta(1+\cos\theta)}\right\}\,.
\end{equation}
The quantity $ F/(4GM/R_1)$ as a function of $ \theta$ is
represented in Fig.2. It has a singularity at the value $
\tan\theta= 0.647799 $ for which the argument of the logarithm in
(\ref{chit}) vanishes. We find $F\approx 1$ for all other values
of $ \theta$ and for all reasonable values of the parameters
involved.
\begin{figure}
\centering
\includegraphics[width=0.6\textwidth]{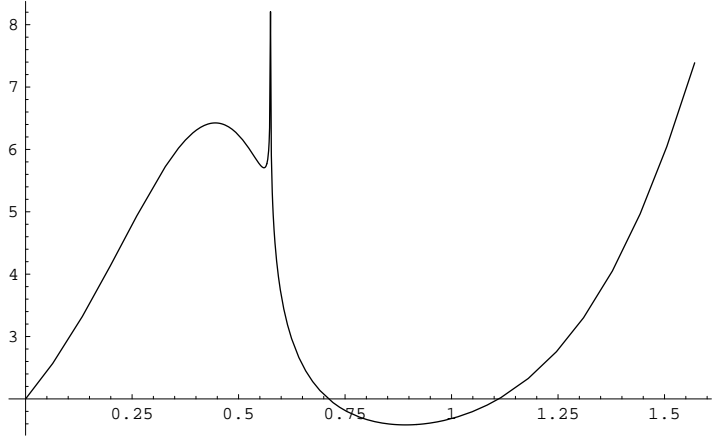}
\caption{\label{fig:CHIT} The plot of $F/(4GM/R_1)$ versus
$\theta$ has a singularity at $ \theta=0.57483$ where the argument
of the logarithm in (\ref{chit}) vanishes. For all other values of
$\theta$ and for reasonable values of $M$ and $R_1$, the
amplification factor $F\approx 1$.}
\end{figure}

\section{\label{sec:5}Conclusions}

Covariant wave equations for massless and massive spin-$1$
particles can be solved exactly to first order in $
\gamma_{\mu\nu}$. The solutions are covariant and invariant with
respect to the gauge transformations of $ a_{\mu}$ and $
\gamma_{\mu\nu}$ and are known when a solution of the free wave
equation is known.

The external gravitational field only appears in the phase of the
wave function. We have shown elsewhere \cite{cai2},\cite{pap2} that
the spin-gravity coupling and Mashhoon's helicity-rotation
interaction \cite{mash},\cite{ryder} follow from $ -i\xi a_{\mu}$
and $ -i\eta a_{\mu}$. According to equations (\ref{10}), (\ref{11})
and (\ref{14}), the spin term $ S_{\alpha\beta}$ finds its origin in
the anti-symmetric part of the space-time connection. In the case of
fermions, $ S_{\alpha\beta}$ is accounted for by the spinorial
connection \cite{punzi}. The term of (\ref{14}) that contains $
S_{\alpha\beta}$ gives rise to the Skrotskii effect.

If the solution of the free wave equation is a plane wave, then
the change in phase is entirely given by $ \Delta\tilde{\varphi}$.
It is from the gravity-induced phase that we have derived the
optics of the particles. The phase can be calculated by means of
path integrations that can be frequently performed in an exact
way, when the metric of the gravitational background is assigned.

From the phases we have derived the geometrical optics of the
particles and verified that their deflection is that predicted by
general relativity. In addition, the background gravitational
field acts as a medium whose index of refraction can be calculated
for any metric from (\ref{gph}), (\ref{photonmomentum}) and $ n =
\tilde{k}/\tilde{k}_0$.

Wave optics can also be extracted from the gravity-induced phases.
In lensing this is normally accomplished by means of a thin lens
model whereby the gravitational potential vanishes everywhere
except on the lens plane. In the present approach the
gravitational field acts all along the particle's trajectory from
source to observer and no thin lens approximation is required.

The exact expression (\ref{FR}) for the phase change $ \Delta
\tilde{\phi}$, gives rise to interference and diffraction
phenomena. In gravitational lensing, wave effects for a point
source depend on the parameter $ \tilde{y}$ and different values
require, in general, different approximations to the solution of
the wave equation. In particular, diffraction effects are expected
to be considerable when $ \tilde{y}\simeq 1$. In the present
approach and in the configuration of Fig. \ref{fig:Lensing},
(\ref{FR}) applies regardless of the value of $ \tilde{y}$ when $
\Delta \tilde{\phi}$ is real.

The extension of our results to include spherical wave solutions
of Helmholtz equation yields an amplification factor (\ref{mf})
whose value can be calculated exactly. We find that essentially
$F\approx 1$ for all reasonable values of the parameters involved.

Because spin has been neglected in the example given, the same
results can also be applied to the case of gravitational lensing
of gravitational waves\cite{tak}.

\section*{Acknowledgments} Research supported by MURST PRIN 2003. The
authors thank E. di Marino for his help in preparing Fig.1.

\appendix

\section{}
\setcounter{section}{1}

In order to prove that (\ref{9}) is a solution of (\ref{6}) and
therefore also of (\ref{4}) when $m=0$, we make use of the formula
\begin{equation}\label{A.1}
  \partial_\tau \int_P^x dz^\lambda G_\lambda (z, x)=G_\tau (x, x)
  +\int_P^x dz^\lambda \partial_\tau G_\lambda(z, x)\,.
\end{equation}

By writing (\ref{9}) in the form
\begin{equation}\label{A.2}
  A_\mu(x)\simeq (1-i\xi(x))a_\mu(x)
\end{equation}
and differentiating $-i\xi a_\mu$ once according to (\ref{A.1}),
we obtain
\begin{equation}\label{A.3}
  \partial_\tau (-i\xi a_\mu)= -\frac{1}{4}\int_P^x dz^\lambda
  \gamma_{[\alpha \lambda, \beta]}(z)
  [\delta_\tau^{[\alpha} \partial^{\beta]} a_\mu(x)
  +K^{[\alpha}\partial_\tau \partial^{\beta]} a_\mu(x)]
 \end{equation}
 \[+\frac{1}{2}\gamma_{\alpha \tau}(x)\partial^\alpha a_\mu(x)+
 \frac{1}{2}\int_P^x dz^\lambda \gamma_{\alpha \lambda}
 (z)\partial_\tau\partial^\alpha a_\mu(x)
 \]
 \[
 + \Gamma_{\sigma,\,\mu\tau}
 (x)a^\sigma (x)+
 \int_P^x dz^\lambda \Gamma_{\sigma,\,\mu\lambda}(z)\partial_\tau a^\sigma(x)\,.
 \]
Differentiating (\ref{A.3}) once more and contracting the indices
of differentiation, we find
\begin{equation}\label{A.4}
\partial^\tau \partial_\tau(-i\xi a_\mu)=
-\frac{1}{2}(\gamma^{\sigma}_{[\sigma,\beta]}\partial^\beta
a_\mu-\frac{1}{2}\int_P^xdz^\lambda\gamma_{[\tau\lambda,
\beta]}\partial^\tau\partial^\beta a_\mu
\end{equation}
 \[
 -\frac{1}{2}\int_P^x dz^\lambda \gamma_{[\alpha\lambda, \beta]}K^\alpha\partial^\tau\partial_\tau\partial^\beta
 a_\mu+\frac{1}{2}\gamma^\tau_{\alpha,\tau}\partial^\alpha
 a_\mu+\gamma_{\alpha\tau}\partial^\tau\partial^\alpha a_\mu
 \]
 \[
 +\frac{1}{2}\int_P^x dz^\lambda
 \gamma_{\alpha\lambda}\partial^\tau\partial_\tau\partial^{\alpha}
 a_\mu+ \partial^{\tau}(\Gamma_{\sigma,\,\mu\tau})a^\sigma
  +2 \Gamma_{\sigma,\,\mu\tau}\partial^\tau a^\sigma+\int_P^x dz^\lambda
 \Gamma_{\sigma,\,\mu\lambda}\partial^\tau \partial_\tau a^\sigma
 \]
We assume that $a_\mu(x)$ is well behaved and that successive
differentiations applied to it commute. Then the second term on the
r.h.s. of (\ref{A.4}) vanishes because two symmetric indices are
contracted on two anti-symmetric indices. The third, sixth and last
terms may be written as $-m^2 (-i\xi)a_\mu$. We find
\begin{equation}\label{A.5}
 \partial^\tau \partial_\tau (-i \xi a_\mu)= -m^2 (-i\xi) a_\mu
 -\frac{1}{2}(\gamma^\sigma_{\sigma, \beta}-2\gamma^\tau_{\beta,
 \tau})\partial^\beta a_\mu+\gamma_{\alpha \tau}\partial^\tau
 \partial^\alpha a_\mu-
\end{equation}
 \[
 +\partial^{\tau}(\Gamma_{\sigma,\,\mu\tau})a^\sigma+ 2
 \Gamma_{\sigma,\,\mu\tau}\partial^\tau a^\sigma\,.
 \]
The second term on the r.h.s. of (\ref{A.5}) may be dropped on
account of (\ref{7}). We now substitute the solution (\ref{A.2})
into the r.h.s. of (\ref{6}) and keep only first order terms. We
get
\begin{equation}\label{A.6}
  \nabla_\nu \nabla^\nu A_\mu +m^2 A_\mu\simeq
  (\eta^{\sigma\alpha}-\gamma^{\sigma\alpha})\partial_\sigma\partial_\alpha
  a_\mu+\partial^\sigma\partial_\sigma (-i\xi)a_\mu
\end{equation}
 \[
 +2 \Gamma_{\sigma,\,\mu\nu}\partial^\nu a^\sigma
 -\frac{1}{2}\gamma_{\sigma\mu, \nu}^{\phantom{\sigma\mu,
 \nu}\nu}a^\sigma+m^2(1-i\xi)a_\mu\,,
 \]
and, on using (\ref{A.5}),
\begin{equation}\label{A.7}
  \nabla_\nu \nabla^\nu A_\mu +m^2 A_\mu\simeq
 -\frac{1}{2}[(\gamma^\sigma_{\sigma, \beta}-2\gamma^\tau_{\beta,
 \tau})\partial^\beta a_\mu +(\gamma_{\mu\tau, \sigma}^{\phantom{\mu\tau,
  \sigma}\tau}-\gamma_{\sigma\tau, \mu}^{\phantom{\sigma\tau,
  \mu}\tau})a^\sigma]=
  \end{equation}
 \[
 = -\frac{1}{2}[(\gamma^\sigma_{\sigma, \beta}-2\gamma^\tau_{\beta,
 \tau})\partial^\beta a_\mu +
 (-\gamma^\tau_{\tau,\mu\sigma}+\gamma^\tau_{\tau,\sigma\mu})a^\sigma]=0
 \]
on account of (\ref{7}).

\section{}

The matrices $ S^{\alpha\beta}$ can be easily obtained from their
definition (\ref{15}) where $ \mu$ and $ \nu$ label, respectively,
rows and columns. We find, for instance,
 \[
 S^{01}=-\frac{i}{2}\left(\begin{array}{cccc}
 0 & 1 & 0 & 0 \vspace{0.05in} \\
 -1 & 0 & 0 & 0 \vspace{0.05in} \\
 0 & 0 & 0 & 0 \vspace{0.05in} \\
 0 & 0 & 0 & 0 \vspace{0.05in} \\
\end{array}\right)\,,
 S^{02}=-\frac{i}{2}\left(\begin{array}{cccc}
 0 & 0 & 1 & 0 \vspace{0.05in} \\
 0 & 0 & 0 & 0 \vspace{0.05in} \\
 -1 & 0 & 0 & 0 \vspace{0.05in} \\
 0 & 0 & 0 & 0 \vspace{0.05in} \\
\end{array}\right)\,.
 \]
Let us define the quantity $A_{\alpha\beta}\equiv
dz^{\lambda}\gamma_{[\alpha\lambda,\beta]} $. On using (\ref{15}),
we obtain
 \begin{equation}\label{B.0}
 A_{\alpha\beta}(S^{\alpha\beta})^{\mu\nu} a_{\nu}=
 \end{equation}
  \[
 -\frac{i}{2}A_{\alpha\beta}\left(\eta^{\alpha\mu}\eta^{\beta\nu}-
 \eta^{\beta\mu}\eta^{\alpha\nu}\right)a_{\nu}=-i\left(A^{\mu0}a_{0}+A^{\mu1}a_{1}+
 A^{\mu2}a_{2}+A^{\mu3}a_{3}\right)\,,
 \]
 which is a four-vector and can also be understood as a
 single-column matrix whose rows are obtained by setting $ \mu =0,1,2,3$.
 On the other hand, using the matrices $ S^{\alpha\beta}$, we also
 find
 \begin{equation}\label{B.00}
 (A_{\alpha\beta}S^{\alpha\beta})(a^{\mu})=
 \end{equation}
  \[
 =2\left[A_{01}S^{01}+A_{02}S^{02}+A_{03}S^{03}+
 A_{12}S^{12}+A_{13}S^{13}+A_{23}S^{23}\right]a^{\mu}=
 \]
\[
 = -i\left(\begin{array}{c}
 A_{01}a^{1}+A_{02}a^{2}+A_{03}a^{3} \vspace{0.05in}\\
 -A_{01}a^{0}+A_{12}a^{2}+A_{13}a^{3}\vspace{0.05in}\\
 -A_{02}a^{0}-A_{12}a^{1}+A_{23}a^{3}\vspace{0.05in}\\
 -A_{03}a^{0}-A_{13}a^{1}-A_{23}a^{2}\vspace{0.05in}\\
 \end{array}\right)\,,\]
which coincides with the final matrix in (\ref{B.0}). Therefore $
a_{\nu}$ is understood as a four-vector in (\ref{B.0}), while in
(\ref{B.00}) $ (a^{\mu})$ is a single-column matrix. This proves the
equivalence of the matrix and tensor expressions used in (\ref{14})
and (\ref{15}). We now apply Stokes theorem to
\begin{equation}\label{B.1}
  A_\mu(x)\simeq a_\mu(x) + \oint_\Gamma dz^\lambda G_\lambda (z, x)
  a_\mu (x)\,,
\end{equation}
where $\Gamma$ is a closed path in Minkowski space. Following
textbook procedures \cite{landau}, we obtain
\begin{equation}\label{B.2}
  A_\mu\simeq a_\mu + \frac{1}{2}\int_\Sigma
  d\tau^{\sigma\delta}\left(\frac{\partial}{\partial z^\delta}G_\sigma-
  \frac{\partial}{\partial z^\sigma}G_\delta\right) a_\mu\,,
\end{equation}
where
\begin{equation}\label{B.3}
  G_\sigma a_\mu = -\frac{1}{2}\gamma_{[\alpha\sigma,
  \beta]}(z)K^\alpha\partial^\beta
  a_\mu(x)+\frac{1}{2}\gamma_{\alpha\sigma}(z)\partial^\alpha
  a_\mu(x)+
\end{equation}
 \[
 +\frac{i}{4}\gamma_{[\alpha\sigma, \beta]}(z)S^{\alpha\beta}a_\mu(x)+\frac{i}{4}\gamma_{\alpha\beta,
 \sigma}(z)T^{\alpha\beta}a_\mu(x)\,,
\]
and $ \Sigma$ is a surface bound by $ \Gamma$. We then find
 \[
 G_{\sigma, \delta}a_\mu=[-\frac{1}{2}\gamma_{[\alpha\sigma,
 \beta]\delta}K^\alpha\partial^\beta+
 \frac{1}{2}\gamma_{[\delta\sigma, \beta]} \partial^\beta+
 \]
 \[
 +\frac{1}{2}\gamma_{\alpha\sigma,
 \delta}\partial^\alpha+\frac{i}{4}\gamma_{[\alpha\sigma, \beta]\delta}
 S^{\alpha\beta}+\frac{i}{4}\gamma_{\alpha\beta,\sigma\delta}T^{\alpha\beta}
 ]a_\mu\,,
 \]
and again
\begin{equation}\label{B.4}
  d\tau^{\sigma\delta}(G_{\sigma, \delta}-G_{\delta,
  \sigma})\frac{a_\mu}{2}=\frac{1}{4}[-(\gamma_{\alpha\sigma,
  \beta\delta}+\gamma_{\beta\delta, \alpha\sigma}-\gamma_{\alpha\delta,
  \beta\sigma}-\gamma_{\beta\sigma,
  \alpha\delta})(x^\alpha-z^\alpha)\partial^\beta
\end{equation}
 \[
 +\frac{i}{2}(\gamma_{\alpha\sigma,
  \beta\delta}+\gamma_{\beta\delta, \alpha\sigma}-\gamma_{\alpha\delta,
  \beta\sigma}-\gamma_{\beta\sigma,
  \alpha\delta})S^{\alpha\beta}]a_\mu d\tau^{\sigma\delta}=
  \]
 \[
 =d\tau^{\sigma\delta}\frac{1}{2}[-R_{\alpha\beta\delta\sigma}(x^\alpha-z^\alpha)\partial^\beta
 +\frac{i}{2}R_{\alpha\beta\delta\sigma}S^{\alpha\beta}]a_\mu\,.
 \]
By introducing the angular momentum generators of the Lorentz
group
\begin{equation}\label{B.5}
  L^{\alpha\beta}=(x^\alpha-z^\alpha)(i\partial^\beta)-(x^\beta-z^\beta)(i\partial^\alpha)
\end{equation}
into (\ref{B.4}) and using (\ref{B.2}), we find
\begin{equation}\label{B.6}
  A_\mu\simeq a_\mu-\frac{i}{4}\int_\Sigma
  d\tau^{\sigma\delta}R_{\alpha\beta\sigma\delta}(L^{\alpha\beta}+S^{\alpha\beta})a_\mu\,,
\end{equation}
which corresponds to (\ref{16}).

\section{}

In order to calculate $\Delta\tilde{\phi}$, it is convenient to
transform all space integrations into integrations over $z^0$. Along
SL we have
\begin{equation}\label{2*}
  U= \frac{-GM}{q_{SL}(z^0)^{1/2}}\,, \quad q_{SL}(z^0)\equiv
  (r_L-z^0)^2+b^{+\, 2}+2(r_L-z^0)b^+ \cos \varphi^+\,,
\end{equation}
 \[
 k^2 = k \cos \varphi^+, \quad k^3= k\sin \varphi^+\,,
 \]
 \[
 \Pi^2 = -i k \cos\varphi^+ +\frac{\cos\varphi^{+}}{r_{L}}\,, \quad \Pi^3 =
 -ik\sin \varphi^+ +\frac{\sin\varphi^{+}}{r_{L}}\,, \quad \mbox{at} \quad z^0=r_L
 \,,\]
where $r_L \sin \varphi^+ = D_{dS}$. We find
\begin{eqnarray}\label{3*}
 -\frac{\Delta \tilde{\phi}_{SL}}{GM} &=&
 \int_0^{r_L}dz^0 q_{SL}(z^0)^{-1/2}[\Pi^0 +\cos \varphi^+
 \Pi^2+ \sin \varphi^+ \Pi^3]+ \\
 &+& 2\int_0^{r_L} dz^0 q_{SL}(z^0)^{-3/2}(-z^0+r_L+b^+
 \cos\varphi^+)(r_L-z^0)\times \nonumber \\
 & & \times
 [-\Pi^2\frac{\sin^2\varphi^{+}}{\cos\varphi^{+}}-\Pi^3\frac{\cos^2\varphi^{+}}
 {\sin\varphi^{+}}+\Pi^0] \nonumber\,,
\end{eqnarray}
where $\Pi^2$ and $\Pi^3$ must take the values specified in
(\ref{2*}) at the upper integration limit. Analogously, for LO we
get
\begin{equation}\label{4*}
  U= -\frac{GM}{q_{LO}(z^0)^{1/2}}\,,
  k^2 = k \sin \theta^+\,, k^3= k\cos \theta^+\,, R_{1}=\sqrt{r_0^{2}+ b^{+2}} \,,
\end{equation}
 \[
 q_{LO}(z^0)\equiv
  (\sqrt{r_0^2+b^{+\, 2}}-z^0+r_L)^2 + r_0^2 -
  \]
  \[
  - 2r_0 (\sqrt{r_0^2+b^{+\, 2}}-z^0+r_L)\cos \theta^+
 \]
 \[
 \Pi^2 = -i k \sin\theta^+ +\frac{\sin\theta^{+}}{R_{1}}\,, \quad \Pi^3 =
 -ik\cos \theta^+ +\frac{\cos\theta^{+}}{R_{1}}\,, \mbox{at} \,\,
 z^0=r_L+R_1
 \]
 and the change in phase is
\begin{eqnarray}\label{5*}
 -\frac{\Delta \tilde{\phi}_{LO}}{GM} &=&
 \int_{r_L}^{r_L+R_1}dz^0 q_{LO}^{-1/2}[\Pi^0 +\sin \theta^+
 \Pi^2+\cos \theta^+ \Pi^3]+ \\
 &+&
 2 \int_{r_L}^{r_L+R_1} dz^0
 q_{LO}(z^0)^{-3/2}(z^0-R_1+r_0
 \cos\theta^+) \times \nonumber \\
 & & \times
 (r_L+R_1-z^0) \left[-\Pi^2\frac{\cos^2\theta^{+}}{\sin\theta^+}-
 \Pi^3\frac{\sin^2\theta^{+}}{\cos\theta^+}+\Pi^0\right]\,.
 \nonumber
\end{eqnarray}
Again, $\Pi^2$ and $\Pi^3$ take the values calculated in (\ref{4*})
at the upper integration limit.

For SP we find
\begin{equation}\label{6*}
  U= -\frac{GM}{q_{SP}(z^0)^{1/2}}\,, \quad q_{SP}(z^0)\equiv
  b^{-\, 2}+(R-z^0)^2-2(R-z^0)b^- \cos \gamma\,,
\end{equation}
 \[
 k^2 = k \cos \gamma, \quad k^3= k\sin \gamma\,, \quad
 \]
 \[
 \cos\gamma=\frac{R^2+b^{-\, 2}-r_s^2}{2b^- R}\,,
 R=\sqrt{D_{dS}+(s+b^-)^2}\,,
 \]
 \[
 \Pi^2 = -i k \cos\gamma +\frac{\cos\gamma}{R}\,, \quad \Pi^3 =
 -ik\sin \gamma +\frac{\sin\gamma}{R}\,, \quad \mbox{at} \quad z^0=R
 \]
and the corresponding change in phase is given by
\begin{eqnarray}\label{7*}
 -\frac{\Delta \tilde{\phi}_{SP}}{GM} &=&\int_0^{R}dz^0 q_{SP}(z^0)^{-1/2}[\Pi^0+\cos \gamma
 \Pi^2+ \sin \gamma \Pi^3] \\
 &+& 2\int_0^{R} dz^0
 q_{SP}(z^0)^{-3/2}(-z^0+R- b^- \cos\gamma)(R-z^0) \times \nonumber \\
 & & \times \left[-\Pi^2\frac{\sin^2\gamma}{\cos\gamma}
 -\Pi^3\frac{\cos^2\gamma}{\sin\gamma} + \Pi^0 \right] \,. \nonumber
\end{eqnarray}
Finally, for PO we obtain
\begin{equation}\label{8*}
  U= -\frac{GM}{q_{PO}(z^0)^{1/2}}\,, \quad  k^2 =- k \sin \theta^-, \quad k^3= k\cos \theta^- \,,
\end{equation}
 \[
q_{PO}(z^0)\equiv r_0^2 + (R_2+R-z^0)^2 - 2r_0(R_2+R-z^0)\cos
\theta^-\,,
 \]
 \[
R_2=\sqrt{r_0^2+b^{-\, 2}}\,,
 \]
 \[
 \Pi^2 = i k \sin\theta^- -\frac{\sin\theta^{-}}{R_{2}}\,, \quad \Pi^3 =
 -ik\cos \theta^- +\frac{\cos\theta^{-}}{R_{2}}\,, \quad \mbox{at} \quad
 z^0=R+R_2
 \]
and the corresponding change in phase is
\begin{eqnarray}\label{9*}
 -\frac{\Delta \tilde{\phi}_{PO}}{GM} &=&\int_{R}^{R+R_2}dz^0 q_{PO}^{-1/2}
 [\Pi^0 -\sin \theta^- \Pi^2+\cos \theta^- \Pi^3] \\
 &+&
 2\int_{R}^{R+R_2} dz^0
 q_{PO}(z^0)^{-3/2}(z^0-R_2-R+r_0
 \cos\theta^-) \times \nonumber \\
 & & \times (R+R_2-z^0)\left[\Pi^2\frac{\cos^2\theta^{-}}{\sin\theta^-}-
 \Pi^3\frac{\sin^2\theta^{-}}{\cos\theta^-}+\Pi^0\right]\,.\nonumber
\end{eqnarray}

\end{document}